\documentclass[JHEP,11pt]{article}
\usepackage{amsmath,amssymb}
\usepackage{booktabs}
\usepackage{epsfig}
\usepackage{graphicx}
\usepackage{color}
\usepackage{shuffle}
\usepackage{wrapfig}
\usepackage{cases}
\usepackage[utf8]{inputenc}
\usepackage{mathtools,tabu}
\usepackage{slashed}
\usepackage[title]{appendix}
\usepackage{setspace}
\usepackage{lineno}

\usepackage{jheppub}
\definecolor{NUpurple}{RGB}{078,042,132}

\newcommand\nothing[1]{}

\author[a]{Zvi Bern,}
\author[b]{Jaroslav Trnka}

\affiliation[a]{Mani L. Bhaumik Institute for Theoretical Physics,\\
UCLA Department of Physics and Astronomy, Los Angeles, CA 90095, USA }
\affiliation[n]{Center for Quantum Mathematics and Physics (QMAP),\\
Department of Physics, University of California, Davis, CA 95616, USA}

\begin{document}

\title{Snowmass TF04 Report: Scattering Amplitudes and their Applications.}

\abstract{The field of scattering amplitudes plays a central role in
  elementary-particle physics.  This includes various problems of
  broader interest for collider physics, gravitational physics, and
  fundamental principles underlying quantum field theory.  We describe
  various applications and theoretical advances pointing
  towards novel descriptions of quantum field theories.  Comments on
  future prospects are included.}

\maketitle

\section*{Executive summary}

Virtually everything we have learned about the behavior of elementary
particles has been gleaned from experimental and theoretical studies
of scattering processes.  The past few decades have taught us that
scattering amplitudes offer remarkable insights into the structure of
quantum field theories, as well as efficient routes to precision
theoretical results needed to interpret modern experiments.  These
insights, including those that follow from novel descriptions of
scattering amplitudes suggest that some of our most cherished notions
about quantum theories using the principles of locality need
revision. These novel approaches include those using on-shell
approaches, twistor-space and geometric approaches.  It has also
become abundantly clear that scattering amplitudes have led to
important progress in issues of interest to the broader community.

Scattering amplitudes have a long history of applications to collider
physics, string theory, supergravity, mathematical physics, and more
recently to gravitational-wave physics, summarized in relevant
Snowmass white papers~\cite{Cordero:2022gsh, LanceWhitePaper,
  Adamo:2022dcm, Buonanno:2022pgc, Goldberger:2022ebt, Shepherd:2022rsg,
  Berkovits:2022ivl,   Bekaert:2022poo, Bourjaily:2022bwx, deRham:2022hpx,
  Kruczenski:2022lot, Giddings:2022jda, Baumann:2022jpr}.  The basic
premise of the field is a virtuous cycle between explicitly
calculating quantities of experimental or theoretical interest and
identifying new structures that teach us basic facts about quantum
field theories.  These structures in turn lead to improved methods to
carry out out new calculations that then lead to new insights.  This
positive feedback loop has continued to infuse the field with new
ideas and energy informing and guiding new advances. This has been
used to push the state of the art for collider physics, supergravity
and more recently for precision calculations of direct importance to
gravitational-wave emission from binary black holes and neutron stars.

Scattering amplitudes also serve as a wonderful theoretical playground
to test new ideas and connections. This has led to great advances in
our understanding of gauge theories, connections to positive geometry
and Amplituhedron, cluster algebras and produced efficient bootstrap
methods for higher-loop amplitudes. We have seen also use of
integrability techniques and an intriguing imprint of AdS/CFT
correspondence in the structure of S-matrix at strong coupling. The
color-kinematics duality and related double copy uncovered a deep
connection between various quantum field theories, and allows us to
construct scattering amplitudes using universal building blocks.

In the coming years we expect that scattering amplitudes will continue
to address nontrivial problems in collider physics, gravitation,
conformal field theories, gravitational-wave physics, 
effective field theories, and as well as open up new direction
such as cosmology and completely unexpected ones.  Tantalizing
hints, such as from new geometric approaches to scattering,
bootstraps and unexplained ultraviolet cancellations in extended
supergravity theories, suggests that we need to rethink
fundamental principles in quantum field theory.


\section{Introduction}
From Rutherford’s discovery of the atomic nucleus more than a century
ago by scattering $\alpha$ particles from gold foil, to the much more
recent discovery of the Higgs boson at the Large Hadron Collider (LHC)
at CERN, the observation and interpretation of scattering events have
been central to our understanding of elementary-particle interactions.
In recent years the field of scattering amplitudes has taken on a
renewed vitality, not only because of the continued importance to
experimental and theoretical studies, but also because of the realization
that scattering amplitudes offer deep insight into the very structure
of quantum field theories. It has had a broad variety of state of the
art applications to collider physics, supergravity, string theory,
mathematical physics and gravitational-wave physics.

The past few decades of research have revealed remarkable new
structures in scattering amplitudes that provide striking insights
into the structure of modern quantum theories, as well as efficient
routes to carry out theoretical results needed to interpret various
experiments.  These insights, including geometric structures in
amplitudes, suggest that some of our most cherished principles about
constructing quantum theories using locality need revision.  It has
also become abundantly clear in recent years that deep issues in
quantum gravity, including its relation to gauge theories, can be
understood through studies of scattering amplitudes.

\begin{figure}[tb]
        \begin{center}
                \includegraphics[scale=.35]{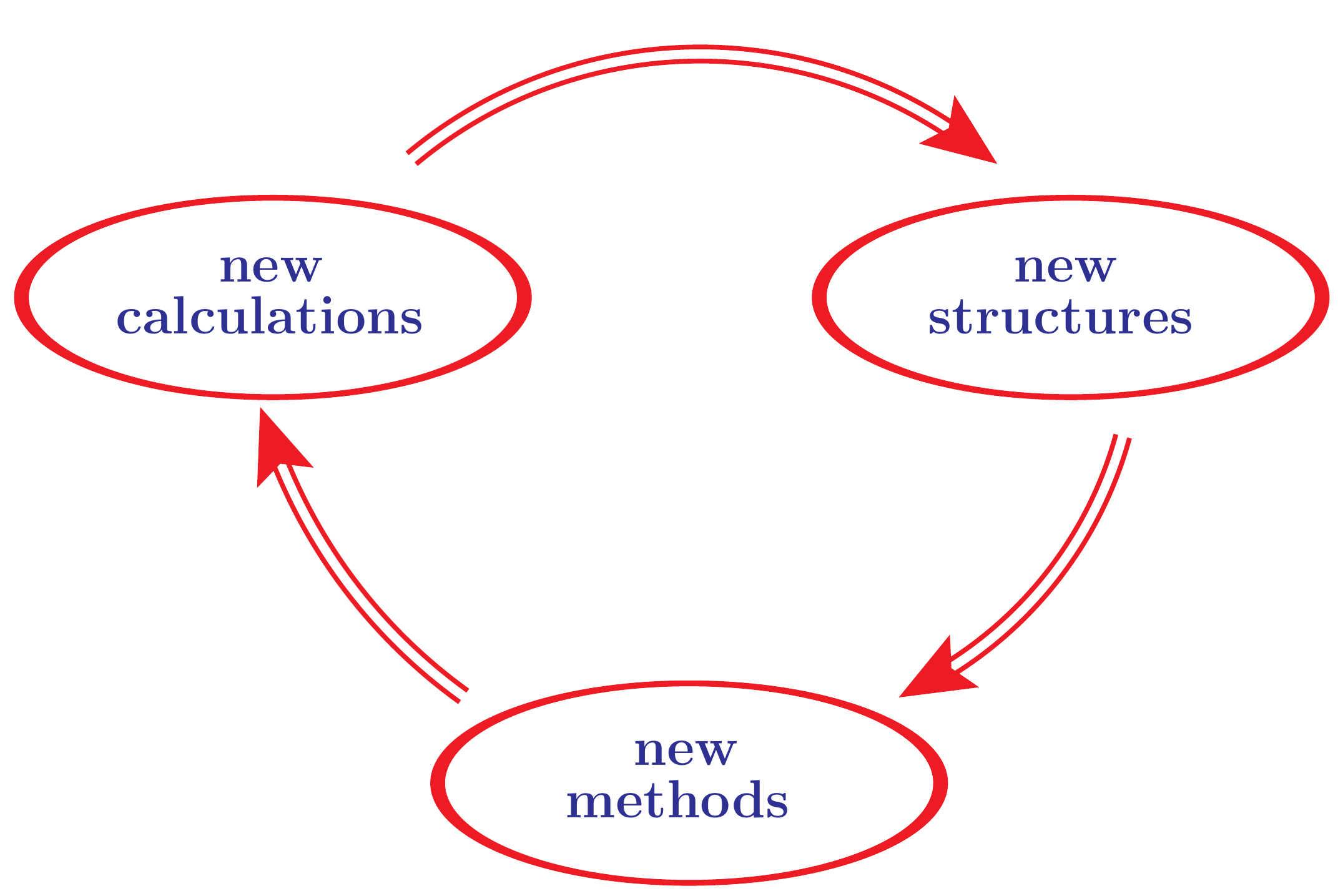}
        \end{center}
        \vskip -.3 cm
        \caption{\small The virtuous cycle between explicit results
          from calculations, new theoretical structures, and new
          methods. }
        \label{fig:Virtuous}
\end{figure}

As illustrated in Fig.~\ref{fig:Virtuous}, a virtuous cycle between
new explicit calculations and new identified structures that then lead
to improved methods is central to progress.  Many examples abound,
starting from the $n$-gluon Parke-Taylor MHV
amplitudes~\cite{Parke:1986gb, Mangano:1987xk}, which was orginally
studied in the context of jet physics at particle colliders.  These
amplitudes form the basis for many other advances including the
construction of $n$-point one-loop MHV amplitudes~\cite{Bern:1994zx},
and the Cachazo--Svrcek--Witten diagrams~\cite{Cachazo:2004kj} for
obtaining all tree-level helicity amplitudes starting from the MHV
ones, as motivated by twistor-space representations of
amplitudes~\cite{Witten:2003nn, Roiban:2004yf}.  The modern unitary
method~\cite{Bern:1994zx,Bern:1994cg,Britto:2004nc} was developed
following computations using earlier methods that led to the simple
form of the explictly computed one-loop five-gluon amplitude of QCD.
The Britto--Cachazo--Feng--Witten (BCFW) on-shell recursion
relations~\cite{Britto:2005fq} were in turn motivated by explicit
forms of tree amplitudes appearing as coefficients of infrared
singularities in one-loop amplitides~\cite{Roiban:2004ix}.  The pace of
development of new methods has continued in recent years with examples
being new methods for describing massive states using helicity
methods~\cite{Arkani-Hamed:2017jhn}, new methods for writing down
multi-loop amplitudes bypassing integration~\cite{Dixon:2015iva} and
improved methods to obtain results directly relevant to precision
predictions of gravitational waves from astrophysical
sources~\cite{Cheung:2018wkq, Kosower:2018adc, Bern:2019nnu}.  There
are many other examples of the synergy between explicit results and
the development of new methods that then lead to further new results,
with the expectation that the cycle will continue well into the
future.

\subsection{Further Reading}
\label{sect:feedback}

Snowmass is a community planning exercise, and the present document
aspires to represent the excitement and interests of the growing
community of theorist who work in the area of scattering amplitudes and
topics with direct overlap.  We gratefully acknowledge the
contributions from the authors of the white papers offering
valuable insights and guidance for the future.  The following people
have contributed to white papers helpful as input for this summary: Tim
Adamo, Nima Arkani-Hamed, Benjamin Basso, Daniel Baumann, Xavier
Bekaert, Nathan Berkovits, Nicolas Boulanger, Jacob L. Bourjaily,
Broedel Broedel, Alessandra Buonanno, Andrea Campoleoni, John Joseph
Carrasco, Mariana Carrillo-Gonz\'alez, Ekta Chaubey, Marco~Chiodaroli,
Claudia de~Rham, Lance J. Dixon, Claude Duhr, Eric D'Hoker, Henriette
Elvang, Fernando Febres Cordero, Dario Francia, Hjalte Frellesvig,
Steven B. Giddings, Walter Goldberger, Daniel Green, Michael B. Green, Maxim Grigoriev,
Martijn Hidding, Henrik Johansson, Austin Joyce, Mohammed Khalil,
Martin Kruczenski, Sandipan Kundu, Robin Marzucca, Andrew J. McLeod,
Tobias Neumann, Donal O'Connell, Enrico Pajer, Joao Penedones,
Guilherme~L. Pimentel, Radu Roiban, Oliver Schlotterer, Ergin Sezgin,
William Shepherd, Evgeny Skvortsov, Mikhail P. Solon, Marcus Spradlin,
Lorenzo Tancredi, Massimo Taronna,Andrew J. Tolley, Jarosalv Trnka,
Matthew Reece, Balt C. van Rees, Charlotte Sleight, Cristian Vergu,
Anastasia Volovich, Matthias Volk, Matt von Hippel, Andreas
von~Manteuffel, Stefan Weinzierl, Matthias Wilhelm, Mao Zeng, Chi
Zhang, and Shuang-Yong Zhou.
 
In this summary, due to the large number of papers, we include
citations only to a relatively small number of selected papers and
refer readers to the relevant white papers~\cite{Cordero:2022gsh,
  LanceWhitePaper, Adamo:2022dcm, Buonanno:2022pgc,Goldberger:2022ebt, Shepherd:2022rsg,
  Berkovits:2022ivl, Bekaert:2022poo, Bourjaily:2022bwx,
  deRham:2022hpx, Kruczenski:2022lot, Giddings:2022jda,
  Baumann:2022jpr} for a detailed list of references.  Besides the
white papers, readers may also consult various review
articles~\cite{Mangano:1990by, Dixon:1996wi, Bern:1996je, Bern:2007dw,
  RoibanReview2011, Carrasco:2015iwa, Cheung:2017pzi, Bern:2019prr, Travaglini:2022uwo}
and books~\cite{Arkani-Hamed:2012zlh, Elvang:2015rqa, Henn:2014yza}
for further details and references.  We also limit our discussion here
to a few selected topics for the purpose of illustrating various
principles as well as the vitality of the field.
A list of pertinent white papers that present an
overview of developments, challenges and new opportunities related to the
field of scattering amplitudes are as follows:

\def\vs{\vskip .2 cm}

\vs\noindent {\bf Computational Challenges for Multi-loop Collider
  Phenomenology}~\cite{Cordero:2022gsh}. High-order precision
computations needed to match the experimental precision at the LHC
 continue to motivate the field of scattering amplitudes
to develop ever more efficient theoretical tools.

\vs\noindent {\bf $\bold{ {\cal N} = 4}$ super-Yang
  Mills}~\cite{LanceWhitePaper}. ${\cal N} = 4$ super-Yang--Mills theory is an important
special case, not only because its relative simplicity compared to QCD
makes it possible to obtain results at spectacularly high-orders, but
it links to both to Maldacena's AdS/CFT conjecture and to supergravity
via the double copy~\cite{Adamo:2022dcm}.

\vs\noindent
{\bf The Double Copy and its Applications}~\cite{Adamo:2022dcm}.
  The double copy began as a means for obtaining gravitational
  scattering amplitudes directly from corresponding gauge-theory
  ones, and has since spread in various directions to a web of
  theories, impacting string theory, particle physics, general
  relativity, and more recently gravitational-wave physics,
  astrophysics, and cosmology.

\vs\noindent {\bf Gravitational Waves and Scattering
  Amplitudes}~\cite{Buonanno:2022pgc}.  Powerful tools from scattering
amplitudes and effective field theory (EFT) have pushed state-of-the-art
perturbative calculations of direct interest to the problem of
gravitational-wave signals from inspiraling binary black holes and
other astrophysical objects.

\vs\noindent {\bf Effective Field Theories of Gravity and Compact
  Binary Dynamics}~\cite{Goldberger:2022ebt}.  The methods of
effective field theory make it possible to diirectly apply scattering
amplitude methods to gravitational-wave physics, with a useful
synergy between the two areas.

\vs\noindent
{\bf Standard Model Effective Field Theory (SMEFT) at the LHC and Beyond}~\cite{Shepherd:2022rsg}.
  The impact of new physics on scattering of Standard Model particles
  can be systematically described by EFTs;  the understanding of
  scattering amplitudes in such EFTs will continue to be an important direction
   in the coming years.

\vs\noindent {\bf String Perturbation
  Theory}~\cite{Berkovits:2022ivl}.  String theory scattering
amplitudes are closely tied to those of quantum field theory and are
an essential part of studies of gravitational physics, dualities and
mathematical structures.  Such studies will continue to lead to new insights.

\vs\noindent {\bf Higher Spin Gravity and Higher Spin
  Symmetry}~\cite{Bekaert:2022poo}.  The problem of consistent
descriptions of higher-spin particles and their scattering amplitudes
continues to be important, and has applications to quantum gravity,
cosmology, conformal field theory, AdS/CFT, string theory, and very
recently to the problem of the coalescence of binary spinning (Kerr)
black holes~\cite{Buonanno:2022pgc}.

\vs\noindent {\bf Functions Beyond Multiple Polylogarithms for
  Precision Collider Physics} \cite{Bourjaily:2022bwx}.  Our ability 
to push the frontiers of scattering amplitudes, whether for collider physics or
more theoretical studies, rely crucially on the mathematics of special
functions.

\vs\noindent {\bf UV Constraints on IR Physics and the S-matrix Bootstrap}~\cite{deRham:2022hpx,Kruczenski:2022lot}. Recent years
has seen a renewed vigor toward fundamental principles, such as
unitarity, crossing and good Regge behavior to constrain low-energy
EFTs, with the goal of identifying the regions
where physically sensible EFTs live.

\vs\noindent {\bf The Deepest Problem: Some Perspectives on Quantum
  Gavity}~\cite{Giddings:2022jda}.  Scattering amplitudes will
continue to be an important tool towards the goal of
realizing a fully satisfactory description of quantum gravity.

\vs\noindent
{\bf The Cosmological Bootstrap}~\cite{Baumann:2022jpr}.  A 
  promising and exciting new direction is to connect the basic principles of scattering
  amplitudes---unitarity, locality and symmetry assumptions--- to the
  study of fluctuations in the early universe.

\section{New Structures from Amplitudes}

Scattering amplitudes display structures with deep implications that
are completely hidden using standard Feynman diagram methods.  The
earliest example of such structures are the
maximally-helicity-violating (MHV) tree amplitudes of quantum
chromodynamics (QCD)~\cite{Parke:1986gb,Mangano:1987xk}.  At the
lowest perturbative (tree-level) order in QCD the $n$-gluon
color-ordered amplitude is given by
\begin{equation}
A(1^-,2^-, 3^+, \ldots, n^+) = i \frac{\langle 1 \,2 \rangle} {\langle
  1 \,2 \rangle \langle 3 \,4 \rangle \cdots \langle n \,1 \rangle}
\,,
\label{ParkeTaylor}
\end{equation}
where the plus and minus signs refer to the helicity of the gluons,
and the notation $\langle i \, j \rangle$ denotes spinor-inner
products. (In this amplitude the color-charges have been stripped
away, giving a color-ordered amplitude from which complete amplitudes
can be reconstructed~\cite{Mangano:1990by,Dixon:1996wi}).  Perhaps the
most remarkable aspect of this formula is its simplicity which may be contrasted
to the complexity of high-multiplicity Feynman diagrams.  The observed
simplicity in Eq.~\eqref{ParkeTaylor} eventually led to the
development of new methods to exploit it, including the unitarity
method~\cite{Bern:1994zx, Bern:1994cg,Britto:2004nc}  and on-shell
recursion~\cite{Britto:2005fq}.  

Over the years the field has continued to identify new and novel
presentations of amplitudes.  Some examples of novel structures
that have driven the development of new methods include,
\begin{itemize}

\item Descriptions of scattering amplitudes in terms of algebraic curves in twistor
space, as motivated by twistor-string theory~\cite{Witten:2003nn, Roiban:2004yf}.

\item Iterative descriptions based on unitarity for loop-level scattering-amplitude integrands
  starting from tree-level ones~\cite{Bern:1994zx, Bern:1994cg, Bern:1997sc, Britto:2004nc}.

\item Recursive descriptions of tree-level scattering amplitudes,
  inspired by the twistor-space description and generalized
  unitarity~\cite{Cachazo:2004kj,Roiban:2004ix,Britto:2005fq}.

\item A hidden ``dual conformal'' symmetry in planar $N=4$ super-Yang-Mills theory~\cite{Drummond:2006rz,Drummond:2008vq} that guides 
 the construction of amplitudes in the theory.

\item Geometric representations of amplitudes, such as the Amplituhedron~\cite{Arkani-Hamed:2013jha},
which seek to recast the basic premise of quantum field theory.

\item A duality between color and kinematics~\cite{Bern:2008qj}, that
  greatly clarifies the double-copy relations between gravity and
  gauge theory, first identified in string theory~\cite{Kawai:1985xq}.

\item New representations of tree amplitudes by an integral over the
  position of $n$ points on a sphere restricted to satisfy a
  set of equations known as ``scattering equations''~\cite{Cachazo:2013hca}.

\end{itemize}
Many of these novel descriptions and structures have led to greatly improved computational methods.

\section{Highlights of Amplitudes}

\begin{figure}[tb]
        \begin{center}
                \includegraphics[scale=.5]{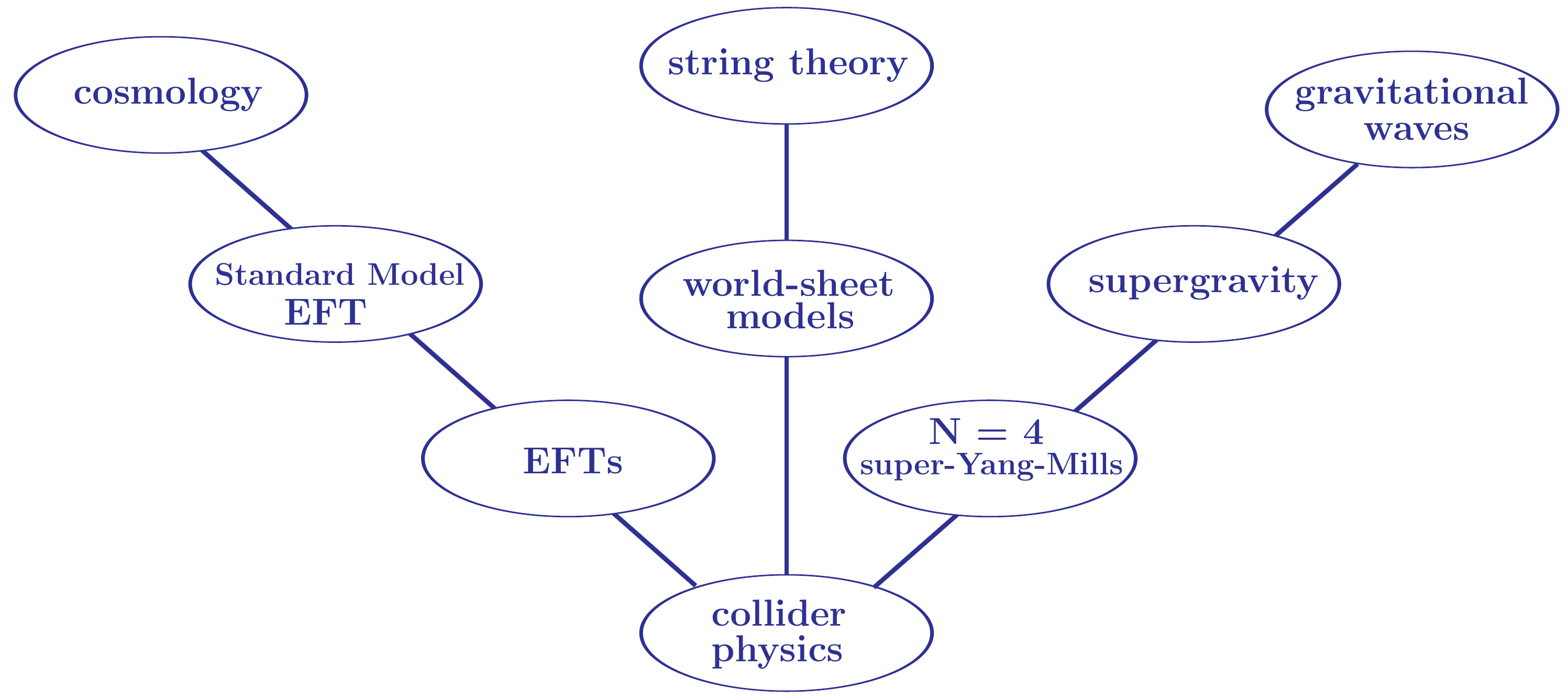}
        \end{center}
        \vskip -.3 cm
        \caption{\small The modern field of scattering  amplitudes began in collider physics 
         and has since been applied to a wide variety of topics. 
       }
        \label{fig:Applications}
\end{figure}

As illustrated in Fig.~\ref{fig:Applications} the modern field of scattering amplitudes has its roots in collider
physics, and has now spread to a large variety of topics.

\subsection{Collider Physics}

The past five years have been a golden era in pushing forward
high-precision calculations in collider
physics~\cite{Cordero:2022gsh}, driven by the unprecedented precision
of the upcoming high-luminosity run at the Large Hadron Collider. Even
rare processes such as Higgs boson production require theoretical
control of cross sections at an unprecedented level of about one
percent. For many processes, current theoretical uncertainties still do
not match the anticipated experimental uncertainties. This continues
to be a primary driver of the field of scattering amplitudes.

In the mid-1980's as collider physics was progressing towards studies
of ever increasing numbers of objects such as jets or vector bosons, as
well as more precise measurements, a need arose for better methods
which became apparent as more calculations were completed.  This cycle
of new collider physics calculations driving new methods continues
today, except now this is occurring at the multi-loop level.

To fully exploit the new-physics discovery potential of on-going
particle-collider experiments, it is essential to continue to reduce
theory uncertainties in order to match experimental improvements. The
current lack of new physics discoveries at the LHC beyond the landmark
discovery of the Higgs boson, emphasizes the importance of the quest
for ever more precise measurements that can tease out long-awaited new
physics. To reach the required level of precision requires control
over all aspects of the collision including parton distributions
functions, the final state parton shower, soft physics associated with
hadronization, as well as the partonic high-energy interactions.

The field of scattering amplitudes is especially attuned to dealing
with the problem of precision interactions at colliders.  Scattering
amplitudes play a central role in virtually all collider phenomenology
predictions and contain the essential dynamical information associated
with the underlying physics.  In order to achieve percent-level
control generally one needs at least two-loop calculations of
scattering amplitudes.  Despite numerous advances in recent years,
such calculations remain challenging, in particular those that depend
on multiple scales associated with either masses or kinematic
invariants. In the past five years many such new calculations have
been performed thanks in part to advances in our understanding of the
analytic structure of scattering amplitudes in quantum field
theory~\cite{Bourjaily:2022bwx}.  These calculations include, for
example, two-loop amplitudes for $2\rightarrow 2$ and $2\rightarrow 3$
processes with multiple scales, three- and four-loop form factors.
Important progress has been accomplished also for the calculation of
related quantities, for example the complete five-loop beta function
and first results for the four-loop splitting functions. (See the
white paper~\cite{Cordero:2022gsh} for references.) A central
issue is to develop ever improved methods for reducing multiloop
Feynman integrals to a small number of master integrals that can be
evaluated using advanced methods.  This progress is not only important
in collider physics but the same advances carry over to other areas
such as to conformal field theories, such as $N = 4$ super-Yang--Mills
theory~\cite{LanceWhitePaper}, and problems in gravitational-wave
physics~\cite{Buonanno:2022pgc} which involve similar integrals.

Besides precision Standard Model calculations another crucial
direction in collider physics is quantifying new physics models.  In
recent years effective field theories have risen in prominence as a
means for systematically catagorizing physics beyond the Standard
Model~\cite{Buchmuller:1985jz,Shepherd:2022rsg}.  Scattering ampltitudes can aid this in
two direction, firstly being useful for organizing the independent
interactions and also by providing tools for computing useful
quantities such as anomalous dimensions and cross-sections. It has also
proven useful for explaining new structures, such as nontrivial
zeroes~\cite{Cheung:2015aba} in anomalous dimension matrices of the
Standard Model Effective Field Theory~\cite{Alonso:2013hga}.

\subsection{Scattering Amplitudes and Gravitational Waves}

Arguably the most exciting recent application of scattering amplitudes
methods is to calculate new quantities of interest to the
gravitational-wave community~\cite{Buonanno:2022pgc}.
The experimental detection of gravitational waves has fundamentally
transformed key areas in astronomy, cosmology, and particle physics,
and will continue to do so for many decades to come given the
anticipated advances in current and future detectors.  This era of
ever-increasing sensitivities holds the promise of dramatic new and
unexpected discoveries, but relies crucially on complementary advances
in our theoretical modeling of gravitational-wave sources.  In recent
years, a new program for understanding the nature of
gravitational-wave sources based on tools from scattering amplitudes
and effective field theory (EFT) has emerged.

At first sight the subject of scattering amplitudes in quantum field
theory might seem to be rather distant from the problem of
gravitational waves.  Firstly the sources are purely classical, then
all the measured events are for bound states not unbounded scattering
processes, and finally the basic object are kilometer scale not
point-like elementary particles.  On the other hand, compared to the
even larger scale of the orbit and the wavelength of gravitational
wave the black holes are effectively points making it possible to use
effective field theory methods~\cite{Goldberger:2004jt,
  Goldberger:2022ebt}.  Given the success of performing high-order
calculations in quantum gravity, even through five loops, one might
suspect that if powerful amplitude methods could be applied to the
gravitational-wave problem one would be able to make progress (see
e.g. Refs.~\cite{Neill:2013wsa, Bjerrum-Bohr:2013bxa, Cheung:2018wkq,
  Bern:2019nnu, Buonanno:2022pgc}).

\begin{figure}[tb]
        \begin{center}
\includegraphics[scale=.45]{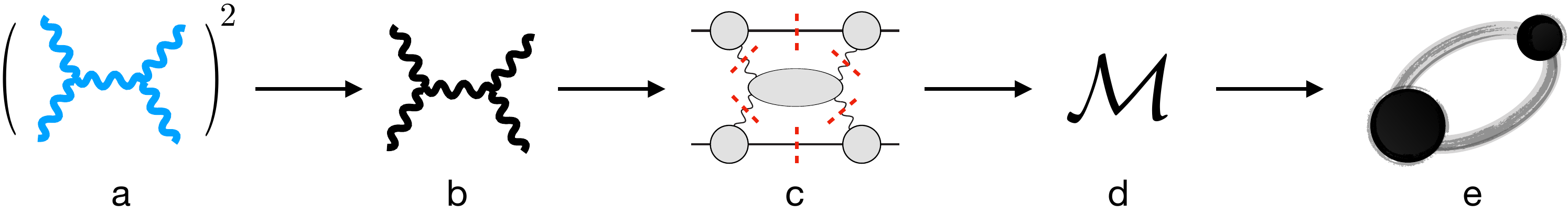}   
        \end{center}
        \vskip -.3 cm
        \caption{\small Amplitudes based methods start from the double
          copy (a) which constructs gravity scattering amplitudes (b)
          from gauge-theory ones. Generalized unitarity (c) then
          builds higher-order loop scattering amplitudes (d) from which
          the interactions (e) between black holes or
          neutron stars can be extracted. From Ref.~\cite{Buonanno:2022pgc}.}
        \label{fig:GravWave}
\end{figure}

The scattering amplitudes community has taken on the challenge,
pushing forward the state of the art in a variety of directions
(summarized in Ref.~\cite{Buonanno:2022pgc}).  A specific request from
the gravitational-wave community~\cite{Damour:2017zjx}
to obtain the conservative two-body
Hamiltonian the third order in Newton's constant
was soon answered~\cite{Bern:2019nnu,Bern:2019crd} and more recently
at the next order as well.  New ideas have been flourishing that link
scattering amplitudes to problems of direct importance (see, for
example, Refs.~\cite{Bini:2021gat, Khalil:2022ylj}) to theorists
working on precision predictions for LIGO/Virgo/KAGRA and future
detectors.  This includes the development of new methods (see
e.g. Refs.~\cite{Cheung:2018wkq, Kosower:2018adc,Bern:2019crd}) for
linking scattering amplitudes to physical observables in the
bound-state gravitational-wave problem. As illustrated in
Fig.~\ref{fig:GravWave}, amplitudes-based methods start with
tree-level scattering which are amenable to double-copy methods, which
then feed into unitarity methods, leading to a scattering amplitude
from which the classical interactions between two black holes or other
astrophysical objects can be extracted.  New progress has also been
accomplished on radiation effects, tidal effect and spin effects.
Higher-spin theories are well studied in particle
physics~\cite{Bekaert:2022poo}, further aiding recent progress.  An
interesting open problem is to apply scattering-amplitude methods to
dissipative effects from the absorption of energy by black holes or
neutron stars.  Based on the recent advances it is clear that
scattering amplitudes will continue to play a prominent role in
pushing forward state-of-the-art perturbative gravitational-wave
computations.

\subsection{Planar ${\cal N}=4$ super-Yang--Mills Amplitudes: from Weak to Strong Coupling}

Maximally supersymmetric Yang--Mills (SYM) theory in the planar limit is from
many perspectives the \emph{simplest quantum field theory}, making it
an ideal toy model for testing amplitude methods that can then be
applied to other theories including QCD. One such success is the
unitarity method which was first developed for maximally
supersymmetric Yang--Mills theory~\cite{Bern:1994zx,Britto:2004nc} and
then later extended to QCD (see e.g.~\cite{Berger:2008sj}).  The
perturbative expansion of the maximally supersymmetric theory has a
number of simplifying features: it is convergent, its scattering
amplitudes are ultraviolet finite and it exhibits a hidden
infinite-dimensional Yangian symmetry~\cite{Bena:2003wd}. The
structure of the scattering amplitudes at weak coupling is
surprisingly simple and follows various organizational principles such
as maximal transcendentality, connections to cluster algebras, and
allow for powerful bootstrap
methods~\cite{Caron-Huot:2016owq, LanceWhitePaper}.  These bootstrap
methods offer hope that one day all multi-loop scattering
amplitudes relevant for collider physics can be directly obtained
bypassing the usual step for first finding an integrand that must then
be laboriously integrated.  The leading IR divergence of the maximally
supersymmetric Yang--Mills S-matrix, related to cusp anomalous
dimension, is known to {\it all} loop orders in this case via
integrability methods~\cite{Beisert:2006ez}. The integrability also
plays a crucial role in the flux-tube methods~\cite{Basso:2013vsa}:
the dual (to scattering amplitudes) null polygon Wilson loop can be
calculated using the operator product expansion (OPE) and a new
decomposition in terms of pentagon transitions related to the dynamics
of flux tubes.  Finally, the strong-coupling
limit is controlled by Maldacena's celebrated AdS/CFT
correspondence~\cite{Maldacena:1997re,Gubser:1998bc} and classical
string configurations in AdS${}_5$~\cite{Alday:2007hr}.  In special
cases, it is even possible to resum the entire perturbative series,
linking weak and strong coupling~\cite{Anastasiou:2003kj,Bern:2005iz}.

\subsection{Gravity as a Double Copy of Gauge Theory}

Perhaps one of the more surprising outcomes from studies of scattering
in gravitational theories is the double copy~\cite{Kawai:1985xq,
  Bern:2008qj,Cachazo:2013hca,Bern:2019prr,Adamo:2022dcm}.  The past
few years have seen a burst of interest in this topic.  At its core,
the double provides a means to calculate amplitudes in one theory
using, as input, amplitudes from two technically simpler theories.
The most prominent case gives gravity scattering amplitudes in terms
of two corresponding gauge-theory ones.  Such relations were
originally discovered in the context of string
theory~\cite{Kawai:1985xq,Berkovits:2022ivl} and have since been
greatly clarified via a ``duality between color and
kinematics''~\cite{Bern:2008qj}, which allows gravity amplitudes to be
generated by gauge-theory ones via a simple replacement of color
factors by kinematic factors.  Here color refers to the usual color
charges of nonabelian gauge theories that describe either strong
or electroweak forces. This duality is effectively a map
between color and kinematic factors that extends to a broad variety of
familiar field and string theories.  It has its origins in
perturbative scattering amplitudes but is currently being systematically
extended to generic classical solutions.  The simplest such example
relates the Coulomb solution in electromagnetism to the Schwarzschild
black hole in Einstein gravity~\cite{Monteiro:2014cda}, with many more
sophisticated constructions available such as the Weyl double copy.

As the simplest example of the double copy, $2\rightarrow 2$ graviton
scattering amplitude at lowest order in Einstein gravity are simply
related Yang--Mills (YM) theory,
\begin{equation}
\mathcal M(1,2,3,4) = \Bigl(\frac{\kappa}{2} \Bigr)^2
    \frac{s t}{u} A(1,2,3,4) \times A(1,2,3,4) \,,
\label{FourPointKLT}
\end{equation}
where $A(1,2,3,4)$ is a color-ordered gauge-theory four-gluon partial
scattering amplitude (related to ordinary amplitudes by appropriately stripping off 
group-theory color factors), $\mathcal M(1,2,3,4)$ is a four-graviton tree
amplitude, $\kappa$ is the gravitational coupling to related to
Newton's constant via $\kappa^2 = 32 \pi^2 G_N$, and $s,t,u$ are the Mandelstam
kinematic invariants.
We can summarize the relation heuristically as
\begin{equation}
\hbox{gravity} \sim (\hbox{gauge theory}) \times (\hbox{gauge theory}) \, .
\label{DoubleCopy}
\end{equation}
In a precise sense the double copy gives us a ``multiplication table''
for converting pairs of gauge theories to gravitational theories.

In the coming years we can expect continued development of the double
copy on its practical and theoretical sides.  In supergravity
besides the quest to understand whether {\it all} supergravity
theories can be expressed as double copies of corresponding gauge
theories, it will be important to carry out new higher-loop computations to
finally understand whether all point-like supergravity theories must,
as lore suggests, necessarily be ultraviolet divergent.  The double
copy has also been used in state-of-the-art calculations of interest
to the gravitational-wave community; it will be important see what
further insights emerge from these studies.  General questions about
the set of all allowed higher-derivative interactions that permit a
double copy also remain.  An important outstanding puzzle is to fully
understand the kinematic algebra behind the duality between color and
kinematics.  While there are many examples where double copies for
classical solutions have been identified it would be important to find
a coherent principle for identifying double-copy mappings for generic
classical solutions.

\subsection{String Scattering Amplitudes and World Sheet Models}

The perturbative S-matrix is a central object not only in quantum
field theory but also in string
theory~\cite{Berkovits:2022ivl}. Superstring perturbation theory is a
rich subject which reveals deep connections between QFT amplitudes,
D-branes, gauge/gravity duality, gravitational waves, black holes,
algebraic geometry, and modular forms. The central objects are
superstring amplitudes, which are primarily considered for massless
gravitons, gauge bosons and their respective supersymmetry
partners. These amplitudes are integrals over the moduli spaces of
Riemann surfaces, and exhibit some remarkable simplicity and
unexpected properties, tightly connected to gauge theory in the
infinite string-tension limit. In recent years, the calculation front
of string perturbation theory was pushed to three loops with many
non-trivial new results. New insights were obtained from string
dualities and AdS/CFT correspondence, and string amplitude were shown
to have some fascinating connections to transcendentality principle
and multiple zeta values. Future targets include lifting technical
obstacles to perform higher-loop calculations of string amplitudes,
explore the rich mathematical structure of Feynman integrals
associated with K3 or Calabi-Yau geometries in the context of
higher-genus amplitudes.

There are also two related exciting directions that have benefited
from the advances in the perturbative string theory~\cite{Berkovits:2022ivl}: ambitwistor
string models~\cite{Mason:2013sva} and Cachazo--He--Yuan (CHY)
formalism~\cite{Cachazo:2013hca}. The latter one expresses
field-theory amplitudes as certain integrals over the worldsheet
localized at the points which satisfy scattering equations. The
integrals are built from simple building blocks and manifest
connections between various field theories, including color-kinematics
duality. The ambitwistor string express the same field-theory
amplitudes as the worldsheet correlators of certain ambitwistor
strings defined by a simple worldsheet action. The correlators reduce
to the formulas in the CHY formalism, making a fascinating link
between strings, worldsheets and field-theory amplitudes.  The
existence of such novel descriptions of field-theory scattering
amplitudes points to new underlying principles in quantum field
theories.

\subsection{Web of Theories}

\begin{figure}[tb]
        \begin{center}
                \includegraphics[scale=.32]{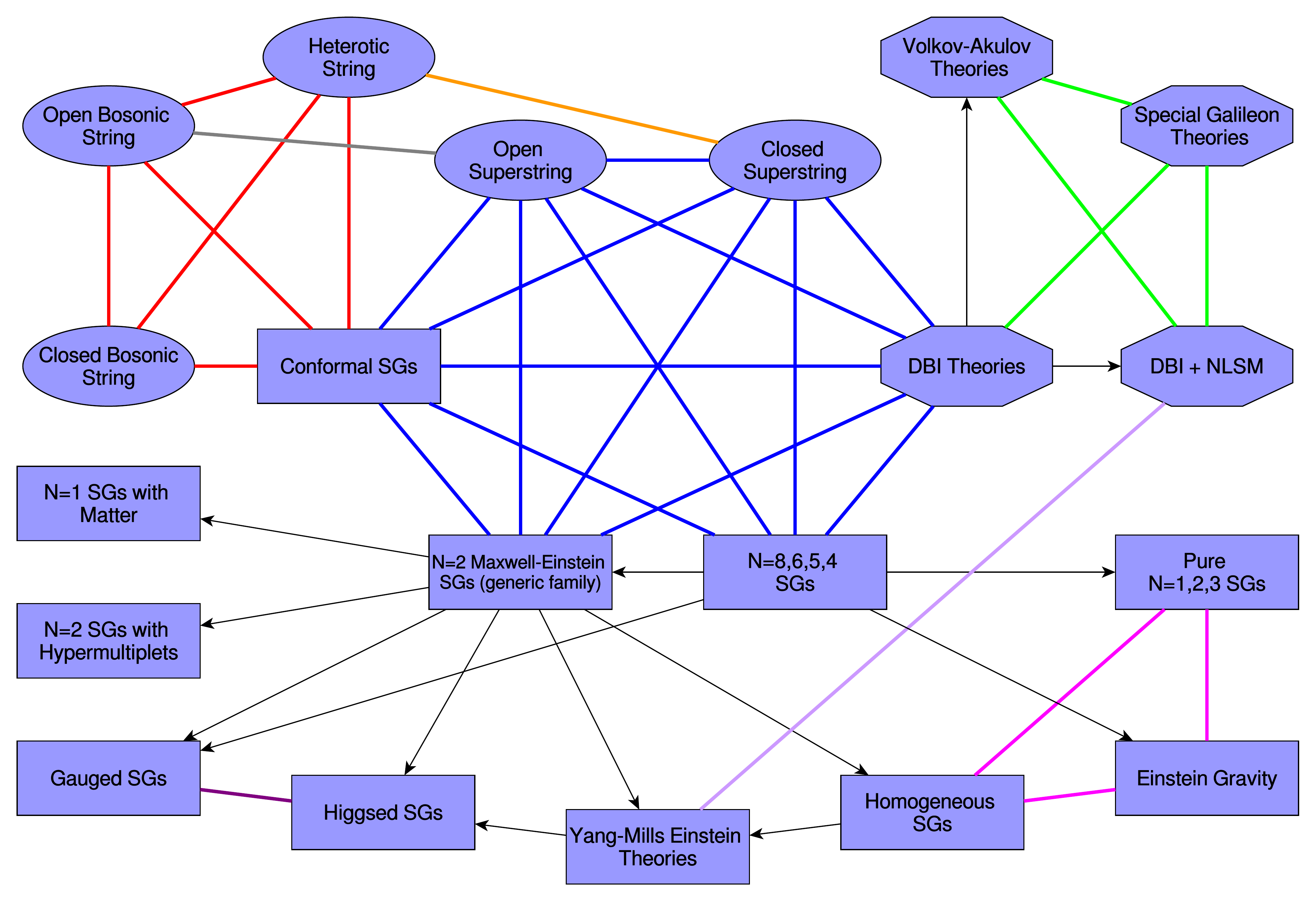}
        \end{center}
        \vskip -.3 cm
   \caption{\small The web of theories exhibited by the double copy,
     as presented in Ref.~\cite{Bern:2019prr}.  The nodes represent
     double-copy-constructible theories, including gravitational
     theories (rectangular nodes), string theories (oval nodes) and
     non-gravitational theories (octagonal nodes). Undirected links
     connect theories with a common gauge-theory factor. Directed
     links are drawn between theories constructed by modifying both
     gauge-theory factors.}
        \label{fig:Web}
\end{figure}

Many scattering-amplitudes developments suggest that there are
nontrivial fundamental links between amplitudes in different
quantum field theories~\cite{Bern:2019prr,Adamo:2022dcm}. Their
tree-level amplitudes are uniquely fixed by simple physical
conditions: locality and unitarity (pole structure and factorizations)
together with additional constraints~\cite{Bern:1994zx,Britto:2005fq}:
gauge invariance in the context of gauge theory and
gravity~\cite{Arkani-Hamed:2016rak}; vanishing soft limits (of various
degrees) for non-linear sigma model (NLSM), DBI action or special
Galileons, Born-Infeld action, and combinations of those for
Volkov-Akulov or Einstein--Yang--Mills actions, and
others~\cite{Cheung:2014dqa,Cheung:2015ota,Cheung:2016drk,Elvang:2018dco}.
The uniqueness of these tree-level amplitudes, for example, allow for
their reconstruction using recursion relations from elementary
amplitudes~\cite{Britto:2005fq}. Infrared physics and soft limits are
also under study from the novel perspective of transforming amplitudes
to celestial sphere~\cite{Cachazo:2014fwa,Pasterski:2016qvg}. In this
picture the soft theorems are understood as symmetries of the
celestial correlators.

The same theories also appear in the context of CHY world-sheet
discription of scattering amplitudes~\cite{Adamo:2022dcm,
  Berkovits:2022ivl, Cachazo:2013hca, Cachazo:2014xea}, and are linked
by double-copy relations: the CHY integrand of the special-Galileon
theory contains two copies of the NLSM integrands, and the same
combination appears in the double-copy construction. In fact,
amplitudes in these special theories can be all built from elementary
building blocks making the connections between theories and special
kinematical behaviors manifest.

A specific example illustrating these ideas are relations between
theories that can be expressed in a double-copy format as a product of
two theories along the lines of Eq.~\eqref{DoubleCopy}.  This type of
multiplication table extends to a remarkably broad variety of theories
beyond standard gravitational theories~\cite{Adamo:2022dcm},
illustrated in Fig.~\ref{fig:Web}, which show links between theories
that share a common theory in the product factor.  A key goal is to
extend these types of relations to a much larger class of theories and
to find new building blocks in a unified description.

\subsection{Constraints on Effective Field Theories}

Physically sensible quantum field theories are constrained by basic
assumptions of unitarity, causality, crossing, and good high-energy
behavior.  The conformal bootstrap program~\cite{Hartman:2022zik} has
emphasized the power of such ideas. These constraints are natural to
apply to scattering amplitudes in the context of effective field
theories that describe physics at scales lower than that of the underlying
fundamental theories~\cite{Adams:2006sv}.  Such effective field
theories are a basic tool for describing physics beyond the Standard
Model~\cite{Shepherd:2022rsg}, including also gravitational theories.

How strongly can we constrain generic EFTs using fundamental
principles that all sensible theories must satisfy? Recent progress
(see
e.g.~Refs.~\cite{Bellazzini:2020cot,Arkani-Hamed:2020blm,Caron-Huot:2021rmr})
systematically constrains the Wilson coefficients of operators or
equivalently the EFT coefficients appearing in amplitudes to bounded
regions. EFTs describing weakly coupled gravity provide an an
important test case, where we can use string theory, as well as
various intermediate energy models to compare the known constraints to
the regions where sensible models actually land~\cite{Bern:2021ppb}.
Remarkably, sensible effective field theories seems to lie on tiny
theory islands far smaller than anticipated from known constraints.
This suggests that new principles constraining physically sensible
effective field theories may very well exist.  An example of such a
principle is the concept of ``low spin dominance''.
Given the recent advances, in the coming years we can
expect much more progress on understanding the islands where
physically sensible EFTs live and identifying new constraints
that all such theories satisfy.

\subsection{Positive Geometry and the Amplituhedron}

A completely new way to define and calculate perturbative scattering
amplitudes has been developed for certain theories as volumes of
positive geometries. The positive geometry encodes the combinatorics
of singularities in the kinematical space and the geometry volume form
reproduces the amplitudes. The prime example is the Amplituhedron
picture~\cite{Arkani-Hamed:2013jha} for planar ${\cal
  N}=4$ SYM theory which defines all tree-level amplitudes and loop
integrands in this theory. Recently the Associahedron geometry has
been linked to amplitudes in scalar $\phi^3$
theory~\cite{Arkani-Hamed:2017mur}. These are new definitions of the
perturbative S-matrix reformulating the physics problem of summing an
infinite number of Feynman diagrams as the mathematical problem of
triangulating geometric spaces. The central object is the positive
geometry~\cite{Arkani-Hamed:2017tmz}, a region in the kinematical
space defined by certain inequalities, and the canonical differential
form on this geometry reproduces the scattering amplitude. This
picture has been used to provide some all-loop order calculations, not
accessible using standard methods~\cite{Arkani-Hamed:2018rsk,
  Arkani-Hamed:2021iya}. There is also a fascinating connection pure
mathematics: Amplituhedron provides a substantial generalization of
the positive Grassmannian and is of great interests to combinatorists;
the positive geometries are closely linked to cluster algebras and
tropical geometries, which are both very active areas of research (see
e.g.~Refs.~\cite{Lukowski:2020dpn,Parisi:2021oql}). The future goals
include uncovering more mathematical connections and using them in the
triangulations and explicit evaluations of differential forms, the
discovery of positive geometries for other quantum field theories, and
formulating a unified geometric picture for the perturbative S-matrix.

This approach stands out in the way it seeks to reformulate
the usual principles of quantum field theories in terms of a
completely different set of geometric principles from which the usual
ones of unitarity, causality and locality follow.  Further progress
offers the promise of radical reinterpretations of quantum field
theory.

\subsection{$S$-Matrix Function Space}

The perturbative scattering amplitudes are kinematical functions of
many variables with special properties dictated by underlying physical
constraints. The poles and branch cuts encode the basic principles of
locality and unitarity, while scaling and various limits encode
universal soft or collinear properties of the $S$-matrix. While
tree-level amplitudes are simple rational functions, the loop
amplitudes are much more complicated and the universe of all functions
that can appear in the loop amplitudes is not well understood even at
next-to-next to leading order in the
coupling~\cite{Bourjaily:2022bwx}, which is necessary to match
experimental precision for many processes~\cite{Cordero:2022gsh}. This
important question has deep mathematical significance---mathematical
properties of these functions can encode hidden physics---as well as
practical use for finding a relatively small basis of objects we need
to consider in any particular calculation. Over the last decade, it
has become clear that a broad range of scattering amplitudes can be
expressed in terms of functions called multiple polylogarithms.  This
realization had led to major computational advances for QCD
amplitudes, as well as remarkably high-loop calculations in
supersymmetric theories.

In planar ${\cal N}=4$ SYM theory~\cite{LanceWhitePaper} knowledge of the function space,
symbols and the connection to cluster
algebras~\cite{Goncharov:2010jf,Golden:2014xqa,Mago:2020kmp}
was used to obtain results for six-point amplitudes up to seven
loops (see e.g.~Refs.~\cite{Dixon:2011nj, Caron-Huot:2016owq,
  Caron-Huot:2019vjl}) for various helicity structures (more precisely
IR finite objects called ``remainder'' and ``ratio'' functions), and
uncover a remarkable new duality which relates the amplitudes to
form factors~\cite{Dixon:2021tdw}.

Unfortunately, scattering amplitudes need further special functions beyond the
polylogarithms, especially in processes with multiple kinematical
variables. This includes multiple polylogarithms, extensions to
elliptic polylogarithms and beyond. Similar functions beyond ordinary
polylogarithms also enter into high-order calculations of
gravitational-wave physics~\cite{Buonanno:2022pgc}.  A summary of the
state of the art, reviewing the ``zoo'' of non-polylogarithmic
integrals and functions and providing future directions is given in
Ref.~\cite{Bourjaily:2022bwx}.

\subsection{Cosmological Bootstrap}

An exciting new direction with important synergies with scattering
amplitudes is the cosmological bootstrap~\cite{Baumann:2022jpr}.  The
physics of primordial density fluctuations is a unique probe of an
early universe. During inflation quantum fluctuations were stretched
to very large distances, and these correlations provide today a rare
insight of the early stages of our universe.  All these are spatial
correlations, and the time only appears in the scale dependence as the
modes freeze at different times depending on the wavelength. The
standard approach to calculate inflationary correlators is to evolve
them from the origin of quantum fluctuations until reheating which
requires the evaluations of complicated time integrals. The
cosmological bootstrap is a new strategy to construct the cosmological
correlation functions using basic physical principles like locality,
unitarity and scale invariance as an approximate symmetry. The grand
goal is to classify all possible patterns of primordial fluctuations
based on these general principles, as well as uncover unexpected
connections between fundamental principles and correlators, theory and
data.  Recent developments in various directions include the study of
constraints imposed by unitarity on cosmological correlations, both in
the perturbative and non-perturbative setup.  These ideas naturally
align with scattering amplitudes so we can expect new fruitful
applications of scattering amplitude methods.

\section{Outlook and Conclusions}

In summary the modern amplitudes program continues to be a vibrant area
significantly impacting various directions including collider physics,
gravitational-wave physics, effective field theory, supersymmetric
gauge and gravity theories, string theory and mathematical physics.
At its core the field of scattering amplitudes is about identifying
new structure that then help us calculate difficult to obtain
quantities of theoretical or experimental interest.  Some of the
identified structures, such as the geometric interpretation of
scattering amplitudes or the double copy, suggest that our basic
description of quantum field theory need revision.  In the coming
years we can expect continuing advances on both the applications
and theoretical sides.

\section*{Acknowledgments}
\label{sect:ack}

We thank the white paper contributors for discussions and for their
insights and efforts.  Z.B. is supported by U.S. Department of Energy
(DOE) under grant No. DE-SC0009937 and thanks the Mani L. Bhaumik
Institute for Theoretical Physics for support.  J.T. is supported by
the DOE grant No. DE-SC0009999 and by the funds of University of
California. 

\bibliographystyle{JHEP}

\bibliography{T4report}

\end{document}